\documentclass[prl,twocolumn,showpacs,superscriptaddress,floatfix]{revtex4}
\usepackage{graphicx,amsfonts,amssymb,amsmath, hyperref}

\newif\ifhyper
\hypertrue
\ifhyper
\hypersetup{
   citecolor = {green},
   colorlinks = {true}, % false
   urlcolor = {blue} % magenta
}
\fi

\newcommand{\beq}{\begin{equation}}
\newcommand{\eeq}{\end{equation}}
\newcommand{\beqa}{\begin{eqnarray}}
\newcommand{\eeqa}{\end{eqnarray}}
\newcommand{\ket} [1] {\vert #1 \rangle}
\newcommand{\bra} [1] {\langle #1 \vert}
\newcommand{\braket}[2]{\langle #1 | #2 \rangle}
\newcommand{\rket} [1] {\vert #1 )}
\newcommand{\rbra} [1] {( #1 \vert}

\begin{document} 

\title{Geometric Entanglement in a One-Dimensional Valence Bond Solid State}

\author{Rom\'an Or\'us}
\email{orus@physics.uq.edu.au}
\affiliation{School of Physical Sciences, The University of Queensland, 
QLD 4072, Australia}

\begin{abstract}

In this paper we provide the analytical derivation of the global geometric entanglement per block for the valence bond solid ground state of the spin-1 AKLT chain. In particular, we show that this quantity saturates exponentially fast to a constant when the sizes of the blocks are sufficiently large. Our result provides the first known example of an analytical calculation of the geometric entanglement for a gapped quantum many-body system in one dimension and far away from a quantum critical point. 

\end{abstract}
\maketitle

{\it Introduction.-} The study of entanglement in quantum many-body systems has attracted great attention in recent years. Quite notoriously, the establishment of a relation between entanglement and the Renormalization Group (RG) has proven particularly fruitful \cite{lat, Zam, fin1, fin2}. In the case of a quantum system in one spatial dimension (1D) with a field theory limit, the $c$-theorem of Zamolodchikov \cite{Zam} involved the existence of a global entanglement loss along successive RG transformations. And not long ago, it was proven that this loss of entanglement is actually fine-grained, in the sense that it can be casted in a set of majorization relations for the reduced density matrices of the ground state of the system \cite{fin1, fin2}. 

Also for quantum systems in 1D, it was recently shown how to quantify the idea of entanglement loss along RG flows in terms of the actual \emph{distance} between the ground state of the system and the closest separable state in the Hilbert space. This novel idea was formulated in terms of the so-called \emph{global geometric entanglement per block}, which was obtained for 1D systems close to and at quantum critical points with an underlying conformal field theory \cite{Bot, Orus}. Despite of its apparent simplicity, the global geometric entanglement has proven very hard to compute. Regarding this, the results from Ref. \cite{Orus} were significant since they managed to obtain analytically this measure of entanglement for a whole class of extended quantum systems as long as they were close to criticality. Nevertheless, the studies of gapped systems in 1D far away from a quantum critical point are almost non-existent and have always been restricted to numerical calculations \cite{GolChain}. Exact and analytic examples of the behavior of the global geometric entanglement far away from 1D quantum critical points are still missing. 

In this paper we address the above situation by computing analytically the \emph{global geometric entanglement per block for the ground state of the AKLT model}. The AKLT model was introduced by Affleck, Kennedy, Lieb and Tasaki \cite{AKLT1, AKLT2}, and constitutes the paradigm of a gapped spin-1 chain with Heisenberg-like interactions supporting Haldane's conjecture \cite{Haldane}. It has also the nice property that its ground state is a one-dimensional Valence Bond Solid (VBS) state which, in fact, is a particular case of a Matrix Product State (MPS) \cite{mps}. Our main result can be stated as follows: in the ground state of the 1D spin-1 AKLT model, the global geometric entanglement per block of size $L$ (which we call $\mathcal{E}(L)$) obeys the law 
\beq
\mathcal{E}(L) = \log{2} - \log{\left[1 + \left(-\frac{1}{3}\right)^L\right] } 
\eeq
in the thermodynamic limit, which saturates at large $L$ to $\mathcal{E}(L \gg 1) \sim \log{2}$.

{\it The AKLT model.-} The AKLT model was introduced by Affleck, Kennedy, Lieb and Tasaki in Refs. \cite{AKLT1, AKLT2} and has become a model of reference in condensed matter physics. Its importance relied originally in that it was the first analytical example of a quantum system that supported Haldane's conjecture \cite{Haldane}, since in the thermodynamic limit it is a local spin-1 Hamiltonian with Heisenberg interactions and a non-vanishing spin gap. More recently, it was shown how to simulate this model with cold atoms in optical lattices \cite{ol}, and its utility for quantum computation was considered in \cite{qc}. Its ground state is a VBS state and, as such, it is closely related to the Laughlin wave function \cite{laughlin} and the fractional quantum Hall effect \cite{corrAKLT}. 

 In the case of a quantum chain of $N$ spins $1$ in the bulk and two spins $1/2$ at the boundaries, the Hamiltonian of the model reads
\beq
H = \sum_{r = 1}^{N-1}\left(\vec{S}^{[r]} \vec{S}^{[r+1]} + \frac{1}{3} (\vec{S}^{[r]} \vec{S}^{[r+1]})^2\right)+ \pi^{[0,1]} + \pi^{[N,N+1]} \ , 
\label{AKLT}
\eeq
where $\vec{S}^{[r]}$ is a vector of spin-1 operators at site $[r]$, and operators $\pi^{[0,1]}$ and $\pi^{[N,N+1]}$ are respectively given by 
\beq
\pi^{[0,1]} = \frac{2}{3}\left(1+\vec{s}^{[0]}\vec{S}^{[1]}\right) \ , ~ 
\pi^{[N,N+1]} = \frac{2}{3}\left(1+\vec{s}^{[N+1]}\vec{S}^{[N]}\right) \ ,
\eeq
where $\vec{s}^{[0]}$ and $\vec{s}^{[N+1]}$ are vectors of spin-1/2 operators at sites $[0]$ and $[N+1]$.
The ground state of the system can be represented in different ways. Here we choose to work with the representation
\beq
\ket{\Psi} = \left(\otimes_{r=1}^N W^{[r,r']}\right) \ket{\Psi^-}^{[0',1]} \ket{\Psi^-}^{[1',2]} \cdots \ket{\Psi^-}^{[N',N+1]} \ , 
\label{vbs}
\eeq 
where $\ket{\Psi^-}^{[r',r+1]} = (\ket{01} - \ket{10})/\sqrt{2}$ is a singlet state for a pair of ancillary spins 1/2 at sites $[r',r+1]$, and $W^{[r,r']}$ is an isometry that projects a state of two spins 1/2 on their symmetric subspace, which describes a state for a spin 1.  The state constructed in this way is a one-dimensional VBS state, and can be proven to be the unique ground state of the Hamiltonian given in Eq. (\ref{AKLT}). Some of the entanglement properties of this state have been previously considered in the literature \cite{entroAKLT, corrAKLT, singleAKLT, S-AKLT, Nusi}. 

\begin{figure}[h]
\includegraphics[width=0.5\textwidth]{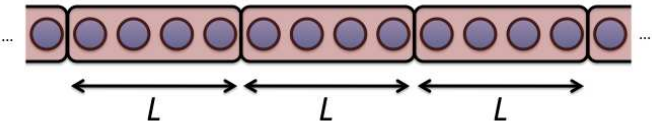}
\caption{(Color online) The 1D spin chain is divided into contiguous blocks of $L$ spins 1. For instance, in this diagram the system has been divided into contiguous blocks of $L=4$ spins.}
\label{fig1}
\end{figure}

{\it Geometric entanglement per block.-} Let us now introduce the measure of entanglement that we shall use as follows. Consider a pure and possibly not normalized quantum state of $N$ parties
$\ket{\Psi} \in \mathcal{H} = \bigotimes_{i = 1}^N \mathcal{H}^{[i]}$, where $\mathcal{H}^{[i]}$ is the Hilbert space of party $i$.  As explained in Ref. \cite{global}, the global multipartite entanglement of $\ket{\Psi}$ can be quantified by considering the maximum fidelity $|\Lambda_{{\rm max}}|$ between the quantum state $\ket{\Psi}$ and all the possible separable and normalized states $\ket{\Phi}$ of the $N$ parties, 
\beq
|\Lambda_{{\rm max}}| = {\rm max}\left|\frac{\braket{\Phi}{\Psi}}{\sqrt{\braket{\Psi}{\Psi}}}\right| \ .
\label{mma}
\eeq
In order to have a well defined measure of entanglement we take the natural logarithm, 
\beq
E(\Psi) = - \log{\left(|\Lambda_{{\rm max}}|^2\right)} \ .
\label{ge1}
\eeq
We will be interested in the above quantity per party, which has a well defined limit when $N \rightarrow \infty$: 
\beq
\mathcal{E}_N = N^{-1} E(\Psi) ~ , ~~~~~~ \mathcal{E} \equiv \lim_{N \rightarrow \infty} \mathcal{E}_N \ . 
\label{ge2}
\eeq
The above quantity has been considered in a variety of situations \cite{Bot, Orus, GolChain, global, LMG}. In this work we will always consider the case in which the one-dimensional quantum system is of infinite size. Moreover, the different parties will be contiguous blocks of $L$ spins 1 as shown in Fig. (\ref{fig1}). The entanglement measure $\mathcal{E}$ corresponds then to the global geometric entanglement per block of size $L$ in the thermodynamic limit \cite{Bot, Orus}.

{\it MPS representation of the ground state.-} For the purpose of this paper, we represent the VBS state in Eq. (\ref{vbs}) in terms of a  MPS. Let us start by writing the singlets in Eq. (\ref{vbs}) as $\ket{\Psi^-} = \sum_{a',a} d_{a',a} \ket{a'a}$, where both indices $a'$ and $a$ can take the values $0$ or $1$, and where $d_{0,1} =  -d_{1,0} = 1/\sqrt{2}$. Also, the isometries $W^{[r,r']}$ in Eq. (\ref{vbs}) are chosen such that they are equal at all sites $W^{[r,r']} = W ~ \forall [r,r']$. These can be written as $W = \sum_{i,a,a'} W^i_{a,a'} \ket{i}\bra{aa'}$, where $i = -1,0,1$ and both $a$ and $a'$ can be either $0$ or $1$. Importantly, in this work we choose $W$ such that 
\beq
WW^{\dagger} = \mathbb{I}_3 \ ,
\label{iso}
\eeq
where $\mathbb{I}_3$ is the identity operator in the three-dimensional Hilbert space of spin 1. This choice of $W$ makes state $\ket{\Psi}$ in Eq. (\ref{vbs}) not to be normalized to one.

\begin{figure}[h]
\includegraphics[width=0.5\textwidth]{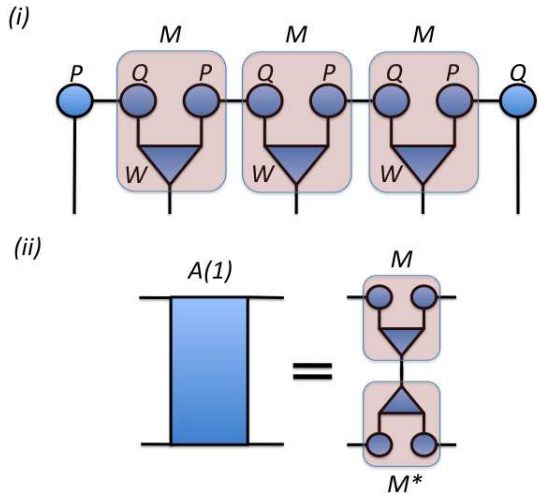}
\caption{(Color online) Diagrammatic representation of (i) an example of MPS decomposition of the VBS state in Eq. (\ref{vbs}) for a system with three spins 1 in the bulk and two spins 1/2 at the boundaries, and 
(ii) MPS transfer matrix $A(1)$ in Eq. (\ref{con}). In these diagrams, tensors are represented by geometric shapes, and indices are represented by emergent legs. A leg shared by two shapes corresponds to the contraction of an index between two tensors, such that there is a sum over all its possible values.}
\label{fig2}
\end{figure}

Let us now write the VBS state as 
\beq
\ket{\Psi} = \sum_{\{a'_0, a_{N+1} ,i\}} c_{a'_0,i_1,\cdots,i_N,j_{N+1}} \ket{a'_0i_1 \cdots i_Na_{N+1}} \ , 
\label{coef}
\eeq
where indices $a'_0 = 0,1$ and $a_{N+1} = 0,1$ represent the states of the spins $1/2$ at the boundaries, and indices $i_r = -1,0,1$ for $r = 1, \cdots, N$ represent the states of the $N$ spins $1$ in the bulk. According to the considerations above, coefficient $c_{a'_0,i_1,\cdots,i_N,a_{N+1}}$ can be decomposed as
\beq
\sum_{\{a_1\cdots a'_N \}} W^{i_1}_{a_1,a'_1}W^{i_2}_{a_2,a'_2} \cdots W^{i_N}_{a_N,a'_N} d_{a'_0,a_1} d_{a'_1,a_2} \cdots  d_{a'_N,a_{N+1}} \ .
\label{mps1}
\eeq
Next, we decompose the coefficients $d_{a',a}$ of the singlets in terms of the product of two matrices $P$ and $Q$ as $d_{a',a} = \sum_{\alpha} P^{a'}_{\alpha}Q^{a}_{\alpha}$, where index $\alpha$ can take the values $0$ or $1$, and where the only non-zero components of $P^{a'}_{\alpha}$ and $Q^{a}_{\alpha}$ are given by
\beqa
P^0_0 = \frac{1}{2^{1/4}}  &,&  P^1_1 = \frac{1}{2^{1/4}} \nonumber \\ 
Q^1_0 = \frac{1}{2^{1/4}} &,& Q^0_1 = -\frac{1}{2^{1/4}} \ .
\label{eq}
\eeqa
Considering together Eqs. (\ref{coef}), (\ref{mps1}) and (\ref{eq}), our final MPS decomposition for the VBS state $\ket{\Psi}$ of Eq. (\ref{vbs}) can be written in terms of matrices $P$ and $Q$ for the spins 1/2 at the left and right boundaries respectively, and a tensor $M$ of components 
\beq
M^i_{\alpha, \beta} = \sum_{a,a'} W^i_{a,a'}Q^a_{\alpha} P^{a'}_{\beta} \ , 
\label{eme}
\eeq
for the spins $1$ in the bulk. This MPS is represented diagrammatically in Fig. (\ref{fig2}.i). 

{\it Geometric entanglement in the ground state.-} As a first step towards the derivation of the global geometric entanglement per block, we compute the MPS transfer matrix $A(1)$ of the ground state of the system. This transfer matrix,  $A(1) = \sum_{\alpha,\beta,\gamma,\delta} A(1)^{(\alpha,\beta)}_{(\gamma,\delta)} \rket{\alpha \beta}\rbra{\gamma \delta}$ (where $\rbra{\cdot} \equiv \rket{\cdot}^T$), has components 
\beq
A(1)^{(\alpha,\beta)}_{(\gamma,\delta)} = \sum_{i} M^i_{\alpha , \gamma} M^{i *}_{\beta , \delta} \ ,
\label{con}
\eeq
which are represented diagrammatically in Fig. (\ref{fig2}.ii). Using Eqs. (\ref{iso}), (\ref{eq}), (\ref{eme})  and (\ref{con}), we obtain the following block-diagonal expression for $A(1)$: 
\beqa
&A(1)& = \frac{1}{2}\left(\rket{11}\rbra{00} + \rket{00}\rbra{11}\right) \nonumber \\
&+& \frac{1}{4}\left( \rket{11}\rbra{11} + \rket{00}\rbra{00} - \rket{10}\rbra{10} - \rket{10}\rbra{01} \right) \nonumber  .
\eeqa
Diagonalizing the above matrix, we get its spectral decomposition 
\beq
A(1) = \frac{3}{4}\rket{e_1}\rbra{e_1} - \frac{1}{4}\left( \rket{e_2}\rbra{e_2} + \rket{e_3}\rbra{e_3} + \rket{e_4}\rbra{e_4} \right) \ ,
\label{spect}
\eeq
where $\rket{e_1} = \frac{1}{\sqrt{2}}\left(\rket{00} + \rket{11}\right)$, $\rket{e_2} = \frac{1}{\sqrt{2}}\left(\rket{00} - \rket{11}\right)$, $\rket{e_3} = \rket{10}$ and $\rket{e_4} = \rket{01}$.

Using the above decomposition we can now easily compute the norm $\langle \Psi | \Psi \rangle$ of the state. For a system with open boundary conditions (OBC), where two spin-$1/2$ particles are attached at the boundaries of the system (sites $[0]$ and $[N+1]$), this norm is given by 
\beq
\braket{\Psi}{\Psi} = \rbra{L}A(1)^N \rket{R} = \left(\frac{3}{4} \right)^N  ~~~ ({\rm OBC}) \ ,
\eeq
where the left and right vectors $\rket{L}$ and $\rket{R}$ are given by $\rket{L} = \rket{R} = \left( \rket{00} + \rket{11}\right)/\sqrt{2}$. 
Similarly, for periodic boundary conditions (PBC) (where all the particles have spin $1$) the norm is given by 
\beq
\braket{\Psi}{\Psi} = {\rm tr} \left( A(1)^N \right) = \left( \frac{3}{4} \right)^N + 3\left(-\frac{1}{4}\right)^N ~~~ ({\rm PBC}) \ , 
\eeq
which matches the result for OBC in the limit $N\rightarrow \infty$, as expected. 

We are now in position to derive the global geometric entanglement per block of size $L$ in the thermodynamic limit. This evaluation follows similar steps to the one in Ref. \cite{Orus}. More precisely, our aim now is to compute the quantity 
\beq
|\Lambda|^2 = \frac{\braket{\Psi}{\Phi} \braket{\Phi}{\Psi}}{\braket{\Psi}{\Psi}} 
\label{la}
\eeq
maximized over all product states $\ket{\Phi}$ of blocks of $L$ contiguous spins $1$. In particular, for $L=1$ we have that $\ket{\Phi}$ is a separable state of all the spins $1$ that define the AKLT model in the thermodynamic limit.

As explained in Ref. \cite{Orus}, the maximization of Eq. (\ref{la}) for a MPS reduces to the problem of finding
\beq
|d|^2 = {\rm max} \left|\rbra{r}^* \rbra{r} A(L)\rket{r} \rket{r}^{*}\right| \ ,
\label{dd}
\eeq
where $A(L)$ is the MPS transfer matrix for blocks of length $L$, and the maximization is done over the vector $\rket{r}$ with the normalization constraint $\rbra{r}^{*} \cdot \rket{r} = 1$. Following Ref. \cite{Orus}, for  $N \gg 1$ we then have that
\beq
|\Lambda|^2 \sim \frac{\left(|d|^2\right)^N}{\braket{\Psi}{\Psi}} \sim  \left(\frac{|d|^2}{(3/4)^L}\right)^N \ ,
\label{ap}
\eeq
where $L$ is the size of the block. 

In order to evaluate Eq. (\ref{dd}), we remember the property that $A(L) = A(1)^L$ (see the discussion around Fig. (3) in Ref. \cite{Orus} for a derivation of this property). Thanks to this fact, together with the spectral decomposition from Eq. (\ref{spect}) for $A(1)$, we can evaluate $A(L)$ for any block size $L$. The maximization in Eq. (\ref{dd}) can then be done in the following way: assume that $\rket{r} = \alpha \rket{0} + \beta \rket{1}$, where $\alpha$ and $\beta$ are complex numbers. Then, we have that 
\beq
\rket{r} \rket{r}^{*} = |\alpha|^2 \rket{00} + |\beta|^2 \rket{11} + \alpha^* \beta \rket{01} + \alpha \beta^* \rket{10} \ .  
\eeq
In terms of $\alpha$ and $\beta$ the quantity $|d|^2$ reads 
\beq
|d|^2 = \frac{1}{2} \left( |\alpha|^2 + |\beta|^2 \right) \left( \left( \frac{3}{4} \right)^L + \left(-\frac{1}{4} \right)^L \right) \ . 
\eeq
Since the vector $\rket{r}$ must be normalized to one, we have that $|\alpha|^2 + |\beta|^2 = 1$, and therefore
\beq
|d|^2 = \frac{1}{2} \left( \left( \frac{3}{4} \right)^L + \left(-\frac{1}{4} \right)^L \right) \ \ \ \forall L  
\label{we}
\eeq
regardless of the values of $\alpha$ and $\beta$. Quite conveniently, no maximization is needed at all in Eq. (\ref{dd}) \footnote{For $L$ even, this implies that $A(L)$ is proportional to a mixed state of two qubits which is equidistant to all bipartite product states.}.

Finally, using Eqs. (\ref{ge1}), (\ref{ge2}), (\ref{ap}) and (\ref{we}) we get our result for the global geometric entanglement per block of size $L$ in the thermodynamic limit, which reads
\beq
\mathcal{E}(L) = \log{2} - \log{\left(1 + \left(-\frac{1}{3}\right)^L\right) } 
\label{ge3}
\eeq
as claimed in the introduction. 

As expected for a gapped quantum many-body system, Eq. (\ref{ge3}) saturates for large block sizes $L$ in $\mathcal{E}(L \rightarrow \infty) = \log{2} \sim 0.693147$. This saturation takes place exponentially fast and, as seen in Table (\ref{tabu}), happens for relatively small values of $L$. 
\begin{table}[h]
\begin{center}
\begin{tabular}{||c||c||||c||c||}
\hline
~~~~~L~~~~~ & ~~~~~$\mathcal{E}(L)$~~~~~ & ~~~~~L~~~~~ &~~~~~  $\mathcal{E}(L)$~~~~~ \\ 
\hline
\hline 
1 & 1.098610 & 8 & 0.692995 \\
\hline 
2 & 0.587787 & 9 & 0.693198\\
\hline 
3 & 0.730888 & 10 & 0.693130\\
\hline
4 & 0.680877 & 11 & 0.693153\\
\hline 
5 & 0.697271 & 12 & 0.693145\\
\hline 
6 & 0.691776 & 13 & 0.693148\\
\hline 
7 & 0.693605 & 14 & 0.693147 \\
\hline 
\end{tabular} 
\end{center}
\caption{Some numerical values of $\mathcal{E}(L)$ for different block sizes $L$. At $L = 14$ we have that $\mathcal{E}(L=14) \sim \log{2} \sim 0.693147$ within the considered accuracy.}
\label{tabu}
\end{table}
Furthermore, note that $\mathcal{E}(L)$ decays to a constant value equally fast as other entanglement measures of the system, such as the entanglement entropy $S(L)$ of a block of size $L$ \cite{entroAKLT}:
\beqa
S(L) &=& 2 + \frac{3}{4}\left(1-\left(-\frac{1}{3}\right)^L\right) \log{\left(1-\left(-\frac{1}{3}\right)^L\right)} \nonumber \\
&-&\frac{1}{4}\left(1+3\left(-\frac{1}{3}\right)^L\right)\log{\left(1+3\left(-\frac{1}{3}\right)^L\right)} \nonumber \ .
\eeqa
Notice also that both $\mathcal{E}(L)$ and $S(L)$ decay as fast as the two-point correlation function for large $L$ \cite{AKLT2, corrAKLT}, $\langle \vec{S}^{[1]} \vec{S}^{[L]} \rangle \sim \left(-\frac{1}{3}\right)^L$. A similar behavior is expected for the ground state of the more general bilinear biquadratic Hamiltonian $H = \sum_r \cos{\theta} \vec{S}^{[r]} \vec{S}^{[r+1]} + \sin{\theta} (\vec{S}^{[r]} \vec{S}^{[r+1]})^2$ (which reduces to Eq. (\ref{AKLT}) for $\theta = 0.1024 \pi$) in the so-called Haldane phase $-0.25 \pi < \theta < 0.25 \pi$ \cite{bibi}.

{\it Conclusions.-} Here we have computed analytically the global geometric entanglement per block for the valence bond solid ground state of the spin-1 AKLT chain in the thermodynamic limit. This quantity saturates to a constant exponentially fast for large block sizes. Our result is the first example of an analytical calculation of the geometric entanglement for a gapped quantum many-body system in 1D far away from a quantum critical point. 

The work in this paper opens the possibility of future analytical studies of the geometric entanglement in other gapped 1D quantum many-body systems. For instance, it would be possible to apply the techniques of this paper to investigate the spin-$S$ AKLT chain \cite{S-AKLT}, the inhomogeneous AKLT model \cite{inhom} and the $SU(n)$ generalized valence bond solid states \cite{sun}. 

{\it Acknowledgements.-} We acknowledge support from the Australian Research Council and The University of Queensland.

{}

\end{document}